\documentclass[twoside]{dis09}
\usepackage[latin1]{inputenc}
\usepackage[dvips]{graphicx,epsfig,color}
\usepackage{wrapfig,rotating}
\usepackage{amssymb,amsmath,array}

\begin{document}
\title{Inclusive Single Hadron Production in Neutral Current Deep-Inelastic Scattering at Next-to-Leading Order}

\author{Carlos Sandoval$^1$
%
%
\vspace{.3cm}\\
%
1- Universit\"at Hamburg - II. Institut f\"ur Theoretische Physik \\
Luruper Chaussee 149, 22761, Hamburg - Germany
%
}

\maketitle

\begin{abstract}
A study of inclusive production of single hadrons with finite transverse
momentum in neutral current deep-inelastic scattering has been carried
out. Cross sections have been calculated using perturbative Quantum
Chromodynamics at next-to-leading order and compared to HERA data.
Predictions for light charged hadron production data were also calculated
for large values of $Q^2$ to show the possible effects of the $Z$-boson
exchange.
\end{abstract}
\section{Introduction}
The perturbative QCD approach for computing hadronic cross sections is based on the parton model picture, in which the cross section for any hard scattering process can be written as a convolution of structure ($f_a(x,Q^2)$) and fragmentation ($D_a(x,Q^2)$) functions of partons (quarks and gluons) and a hard-cross section factor~\cite{Catani:1996vz}. The structure and fragmentation functions are non-perturbative, universal quantities, while the hard-cross section can be calculated within perturbative QCD to the lowest order in the running coupling $\alpha_s(Q)$ as long as $Q\gg\Lambda$, where $\Lambda$ is the QCD scale.\\
This naive parton model corresponds to the leading order (LO) approximation. However, due to the perturbative nature of $\alpha_s$, the running of the coupling constant could be hidden in higher order corrections and therefore, the LO calculation can only predict the order of magnitude of a given cross section. The accuracy of the perturbative QCD expansion is then controlled by the size of the higher order contributions. Any perturbative QCD prediction needs then, next-to-leading (NLO) corrections and NLO definitions of the running coupling constant, and the structure and fragmentation functions.\\ 
In this work we have carried out a NLO calculation of single hadron production in deep-inelastic scattering (DIS), using standard PDF sets and fragmentation functions fitted from $e^+ e^-$ data. 
\section{Theoretical formalism}
We are interested in single production of hadrons in DIS: 
\begin{equation}
l(k)+p(P)\to l(k')+ h(p_h) + X,
\end{equation}
where the lepton can be an electron, muon or an (anti)neutrino, and the exchanged vector boson a photon, $W^\pm$ or $Z$. In this work we concentrate only on the neutral current case. The cross section can be written as a convolution of the partonic cross section with the parton distribution functions and the fragmentation functions~\cite{Furmanski:1981cw}:
\begin{eqnarray}\label{mainxsec}
\frac{d \sigma^{h}}{d x dy d z d\phi } &=& \int_{x}^1 \frac{d\bar x}{\bar x}\int_{z}^1\frac{d\bar z}{\bar z} \sum _{ab} f_a^p\left(\frac {x}{\bar x}, \mu^2\right)\frac{d \sigma^{a b}}{d\bar x dy d\bar z d\phi } D_b^h \left(\frac{z}{\bar z},\mu^2\right),
\end{eqnarray}
where we have used the kinematic variables
\begin{eqnarray}
x=\frac{Q^2}{2P\cdot q},\qquad y&=&\frac{P\cdot q}{P\cdot k}, \qquad z=\frac{P\cdot p_h}{P\cdot q} \nonumber \\
\bar x=\frac{Q^2}{2p_a\cdot q},\qquad \bar y &=& y, \qquad \bar z=\frac{p_a\cdot p_b}{p_a\cdot q}.
\end{eqnarray}
The partonic cross section may be split into leptonic and hadronic parts,
\begin{equation}
\frac{d \sigma^{a b}}{dx dy dz  }=\frac{\alpha^2}{16\pi^2}\frac{y}{Q^4}\lambda_{ab} L^{\mu\nu}H_{\mu\nu}^{ab} ,
\end{equation}
where 
\begin{equation}
L^{\mu\nu}=\frac{Q^2}{2y}\left(\frac{2-2y+y^2}{y}\right)\left(-g^{\mu\nu}\right)+\frac{2Q^4}{s_h^2}\left(\frac{y^2-6y+6}{y^4}\right)p_a^\mu p_a^\nu\pm i \frac{Q^2}{s_h}\left(\frac{y-2}{y^2}\right)\varepsilon^{\mu\nu\alpha\beta}p_{a\alpha}q_{\beta}
\end{equation}
with $s_h=\frac{Q^2}{\bar x y}$. The coefficients $\lambda_{ab}$ contain the information about how the fermions couple to the vector bosons:
\begin{eqnarray}
\lambda_{ab}^{T,L}&=&e_f^2-2e_fv_fv_e\chi_Z(Q^2)+\left(a_f^2+v_f^2\right)\left(a_e^2+v_e^2\right)\chi_Z^2(Q^2)\\
\lambda_{ab}^{A}&=&-2e_fa_fa_e\chi_Z(Q^2)+4a_fa_ev_fv_e\chi_Z^2(Q^2),
\end{eqnarray}
where
\begin{eqnarray}
a_{e,f}&=&T_{e,f}^3,\\
v_{e,f}&=&T_{e,f}^3-2e_{e,f} \sin^2\theta_W,\\
\chi_Z(Q^2)&=&\frac{1}{4\sin^2\theta_W\cos^2\theta_W}\frac{Q^2}{Q^2+M_Z^2}.
\end{eqnarray}
$T_{e,f}^3$ is the weak isospin and $\theta_W$ the weak mixing angle. 
\section{Comparisons with HERA data}
\begin{wrapfigure}{r}{0.5\columnwidth}
\centerline{\includegraphics[width=0.45\columnwidth]{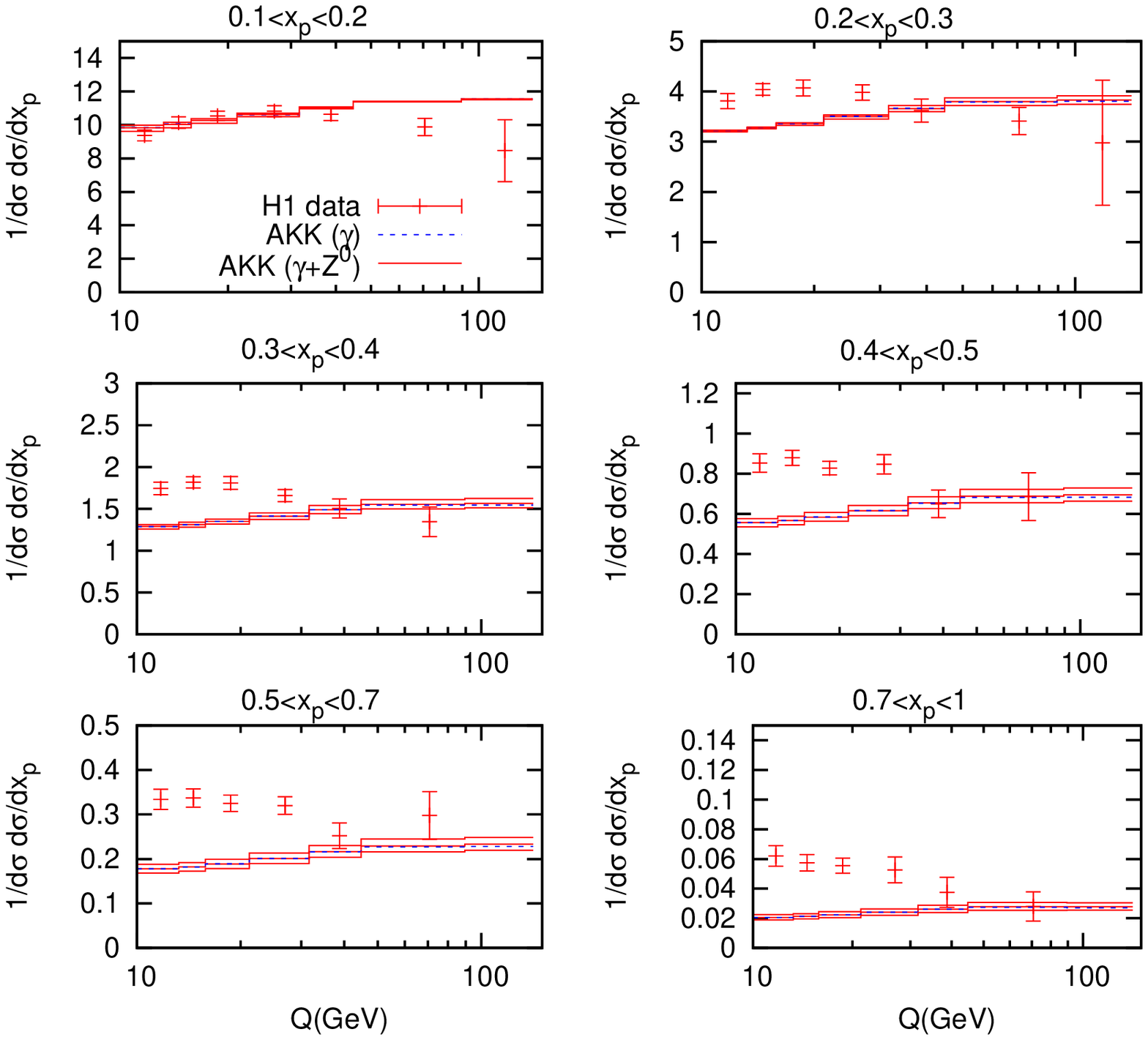}}
\caption{Comparison of theoretical predictions using AKK with H1 data.}\label{h1new}
\centerline{
\includegraphics[width=0.35\columnwidth, angle=270]{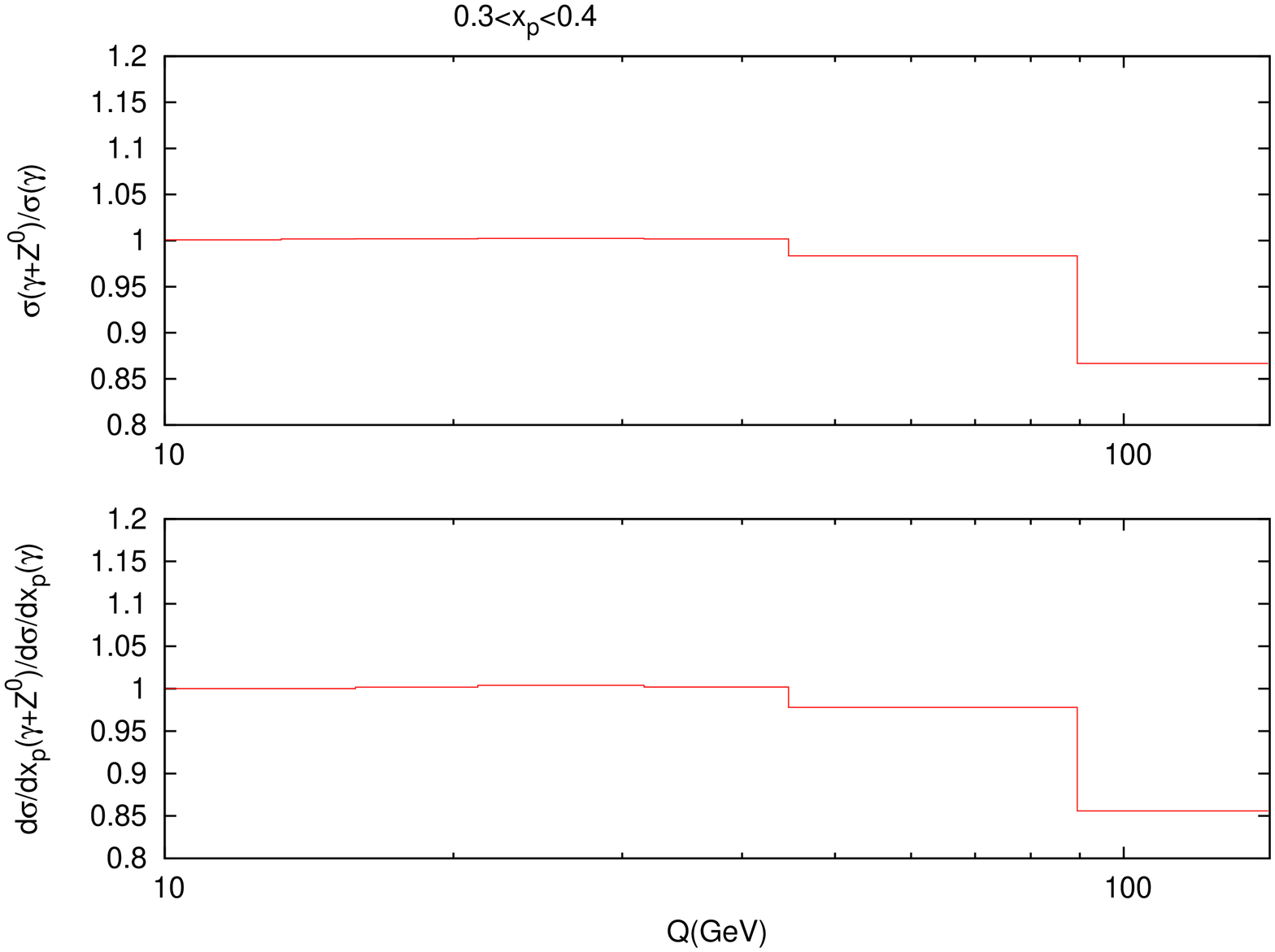}
}
\caption{Ratio between cross sections with and without $Z$ boson contribution to first order in $\alpha_s$.}\label{h1newratio}
\end{wrapfigure}

\begin{figure}[h]
\begin{center}
\includegraphics[width=3.4cm, angle=270]{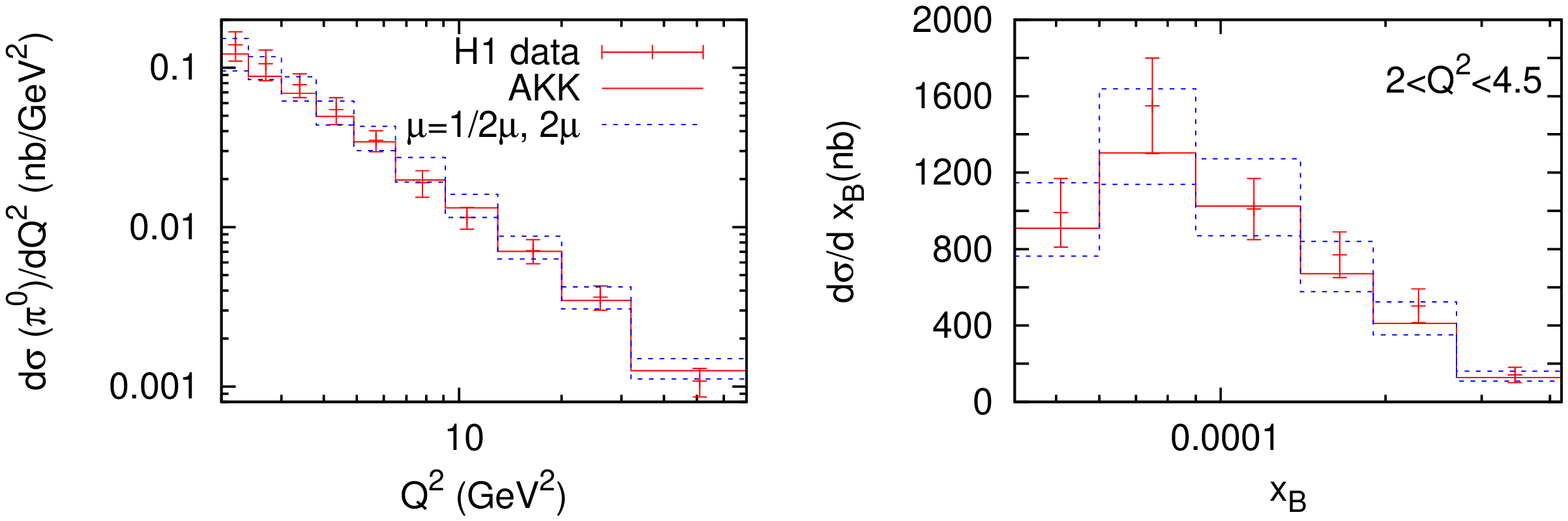}
\caption{Cross section differential in $Q^2$ and $x_B$ for $\pi^0$ production.}\label{pion}
\includegraphics[width=3.4cm, angle=270]{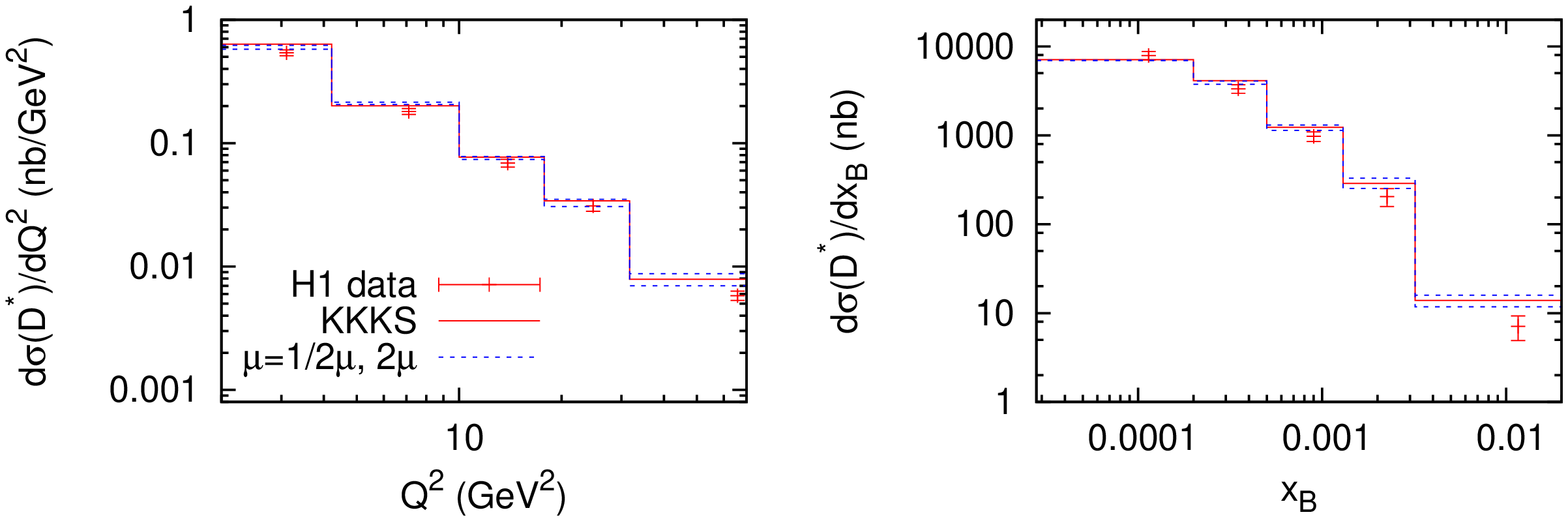}
\caption{Cross section differential in $Q^2$ and $x_B$ for $D^{*\pm}$ production.}\label{dstar}
\includegraphics[width=3.4cm, angle=270]{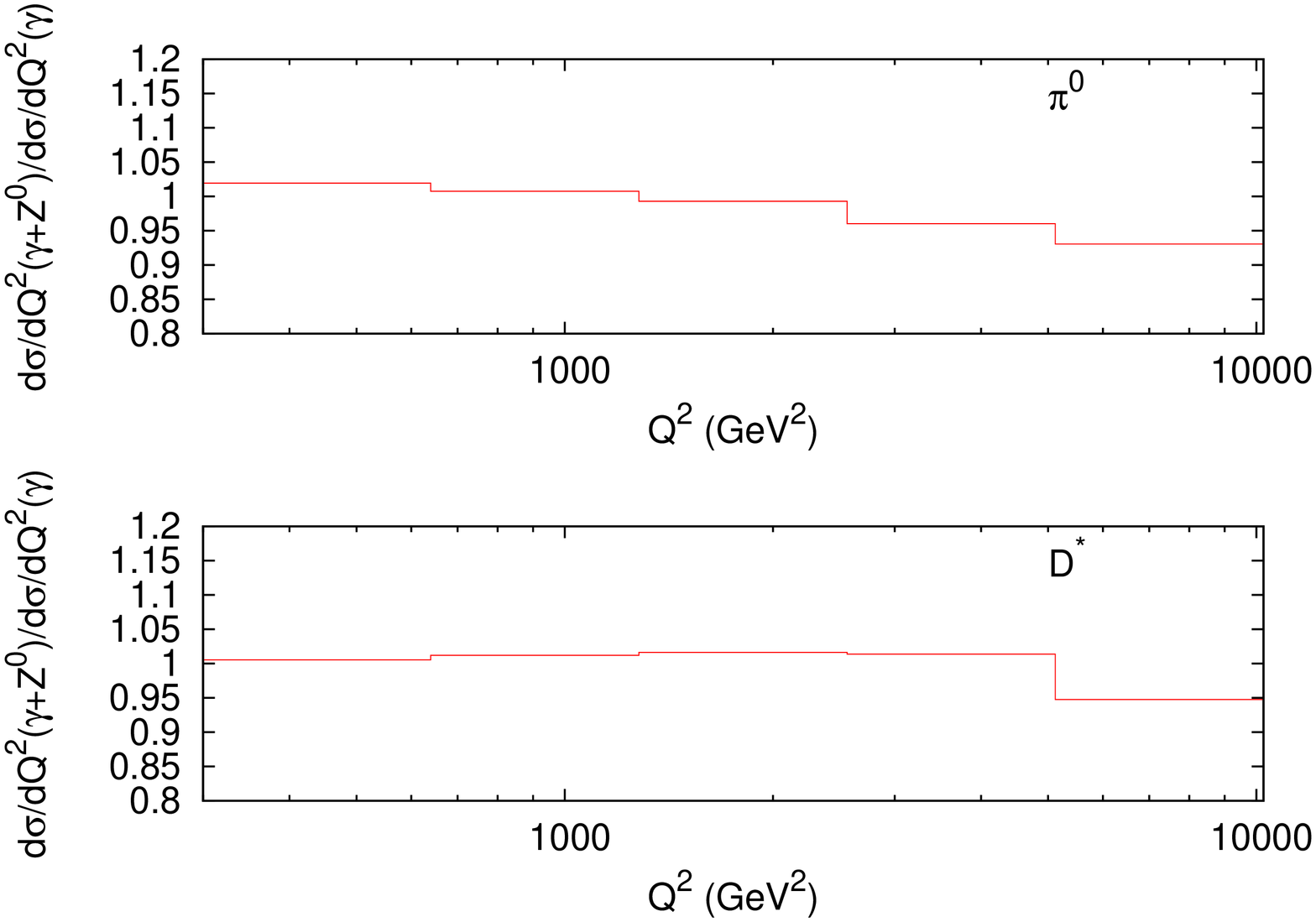}
\caption{Ratio between cross sections with and without $Z$ boson contribution to second order in $\alpha_s$.}\label{ratio}
\end{center}
\end{figure}

In this section we present our numerical results for single hadron inclusive production measurements from HERA data. Cross sections are calculated to NLO in the $\bar{\text{MS}}$ scheme. We set the number of active quark flavours $n_f=5$. We used the CTEQ6.6M PDF set of ref.~\cite{Nadolsky:2008zw} and their value $\Lambda^{(5)}_{\text{QCD}}=226$ MeV.\\
At first order in $\alpha_S$, the only partonic subprocess contributing to LO is $\gamma^*,Z^0+q\to q$. In order to calculate the cross section at NLO we must include virtual and real corrections to this process, and using dimensional regularization and standard renormalization techniques, cancel all singularities that there appear. In this case the factorization/renormalization scale was chosen as $\mu=Q$.\\ New charged particle production data was released in 2007 by the H1 collaboration~\cite{Aaron:2007ds}, that we can use for comparison. There, the cross section differential in the scaled momentum $x_p=\frac{2p_h}{\sqrt{s}}$, normalized to the total cross section, was measured for different bins in $x_p$ and for $100<Q^2<10000$. Our results are shown in Figure~\ref{h1new}, where the solid central line is our main prediction, including the $Z$ boson contribution, and the dashed line is the prediction with only virtual photon exchange. We have used the AKK set of fragmentation functions to make our predictions~\cite{Albino:2008fy}. The two solid lines at the top and bottom of the central line are the predictions obtained when the scale was changed to $\mu=\frac 12 \mu$ and $\mu=2\mu$ respectively, which gives us an idea of the theoretical error in our calculation.\\
Unfortunately, although for these values of $Q^2$, we should be able to see an effect due to the presence of the $Z$ boson, since these cross sections are normalized to the total cross section, and the effect is present in both, they cancel each other out and we can not distinguish between the curve with and without the $Z$ boson contribution. What we can do is plot the ratio of the unnormalized cross sections and there we should be able to see what kind of effect the $Z$ boson introduces. The ratio for the total and the differential cross sections is shown in Figure~\ref{h1newratio}.   
The effect of the $Z$ boson contribution is found to be of about 15$\%$ in this case.\\\\ At second order in $\alpha_s$, the LO contributions are
\begin{eqnarray}
\gamma^*,Z^0+q\to q+g,\nonumber \\
\gamma^*,Z^0+q\to g+q,\\
\gamma^*,Z^0+g\to q+\bar q,\nonumber 
\end{eqnarray}
and we should again calculate virtual and real corrections to these processes. In this case, the cancellation is not straightforward as in the previous case and a suitable method must be used in order to cancel all singularities. We have chosen the dipole subtraction method~\cite{Catani:1996vz}. After carrying out the renormalization procedure for the virtual corrections, all the ultraviolet singularities should cancel and we have only soft and collinear divergences left. The idea of the subtraction method is to use the identity
\begin{equation}\label{sigmanlo}
\sigma^{NLO}=\int_{m+1}(d\sigma^R - d\sigma^A)+\int_{m+1} d\sigma^A + \int_m d\sigma^V,
\end{equation}
where the approximated cross section $d\sigma^A$ is such that is has the same singular behaviour in $d$ dimensions as $d\sigma^R$. In this way, $d\sigma^A$ acts effectively as a counter term for $d\sigma^R$ and introducing the phase space integration, we can safely perform the limit $\epsilon\to 0$ under the integral sign in the first term on the right-hand side of equation~\ref{sigmanlo} and therefore, perform the integral in four dimensions. All the remaining singularities are at this point associated with the last two terms in equation~\ref{sigmanlo}. If we then integrate $d\sigma^A$ analytically over the one parton subspace leading to the $\epsilon$ poles, we can combine these poles with the ones coming from the virtual corrections and in this way, cancel all the singularities. Then we can perform the limit $\epsilon\to 0$ and integrate numerically the rest over the $m$-parton phase space. The final structure of the calculation is 
\begin{equation}
\sigma^{NLO}=\int_{m+1}\left[\left(d\sigma^R\right)_{\epsilon=0} - \left(d\sigma^A\right)_{\epsilon=0}\right]+\int_{m} \left[d\sigma^V + \int_1 d\sigma^A \right]_{\epsilon=0}.
\end{equation}
Using this formalism, we can then obtain predictions for single hadron production at NLO at second order in $\alpha_s$. However, since we are interested in perturbative QCD effects, we require the hadron to carry non-zero transverse momentum ($p_T^*$) in the centre-of-mass frame of the vector boson and the incoming proton. Two measurements are useful for comparison, both from the H1 collaboration (~\cite{Aktas:2004rb} and ~\cite{Aktas:2006py}). The first one is $\pi^0$ production, where we have used again the AKK set of fragmentation functions for our predictions and $p_T^*>2.5$ GeV. For the second one, $D^{*\pm}$ production, we have used the KKKS set of fragmentation functions~\cite{Kneesch:2007ey} and $p_T^*>2.0$ GeV. More details about the kinematic region used in each of these measurements can be found in the experimental papers respectively. Our results are shown in Figures~\ref{pion} and~\ref{dstar}.   
Here again the solid red line represents our prediction and the dashed lines at the top and bottom of it are the corresponding predictions with a different scale as in the previous case. It is not possible to appreciate any effect of the $Z$ boson contribution, since $Q^2$ only goes up to 70 GeV$^2$. However, we can imagine we extend the range of $Q^2$ for these two cases, and as before, we calculate the ratio between the cross sections with and without the $Z$ boson. Our results are shown in Figure~\ref{ratio}, where we can see contributions of up to 7$\%$ for $Q^2>10000$. 
\section{Conclusions}
The NLO cross section for single hadron production was calculated in perturbative QCD for neutral current DIS using the dipole subtraction method. Results agree with the data for $\pi^0$ and $D^{*\pm}$ production using AKK and KKKS fragmentation functions. Effect of $Z^0$ boson in the cross sections found to be up to $15\%$ in the first order calculation and $7\%$ second order calculation.
\section{Acknowledgments}
This work was done in collaboration with Bernd Kniehl and Gustav Kramer. 
\section{Bibliography}

\begin{footnotesize}

\bibliographystyle{unsrt}
\bibliography{sandoval_carlos}
%

\end{footnotesize}

\end{document}